\documentclass[letterpaper, 10 pt, conference]{ieeeconf}  
\usepackage{amsmath,graphicx}
\usepackage{amsfonts}
\usepackage{algorithmic}
\usepackage{graphicx}
\usepackage{textcomp}
\usepackage{booktabs}
\usepackage{tabularx}
\usepackage{rotating}
\usepackage{adjustbox}
\usepackage{amsmath}
\usepackage{eucal}
\let\labelindent\relax
\usepackage{enumitem}

\IEEEoverridecommandlockouts                              

\title{\LARGE \bf
Multi-Contrast MRI Segmentation Trained on Synthetic Images
}

\author{Ismail Irmakci$^{1}$, Zeki Emre Unel$^{2}$, Nazli Ikizler-Cinbis$^{2}$ and Ulas Bagci$^{1}$
\thanks{$^{1}$Machine and Hybrid Intelligence Lab, Northwestern University Feinberg School of Medicine, Chicago, IL 60611, USA}
\thanks{$^{2}$Department of Computer Engineering, Hacettepe University, Ankara, Turkey
        {\tt\small}}%
\thanks{$^{*}$Correspondence to ulas.bagci@northwestern.edu
         {\tt\small}}%
}

\begin{document}

\maketitle
\thispagestyle{empty}
\pagestyle{empty}

\begin{abstract}
In our comprehensive experiments and evaluations, we show that it is possible to  generate multiple contrast (even all synthetically) and use synthetically generated images to train an image segmentation engine. We showed promising segmentation results tested on real multi-contrast MRI scans when delineating muscle, fat, bone and bone marrow, all trained on synthetic images. Based on synthetic image training, our segmentation results were as high as 93.91\%, 94.11\%, 91.63\%, 95.33\%, for muscle, fat, bone, and bone marrow delineation, respectively. Results were not significantly different from the ones obtained when real images were used for segmentation training: 94.68\%, 94.67\%, 95.91\%, and 96.82\%, respectively.
\newline

\indent \textit{Clinical relevance}—  Synthetically generated images could potentially be used in large-scale training of deep networks for segmentation purpose. Small data set problem of many clinical imaging problems can potentially be addressed with the proposed algorithm. 
\end{abstract}

\section{INTRODUCTION}

Segmentation of medical images is a long-standing problem with excessive number of deep learning based methods already available nowadays. Although there are recent paradigm-shifting works at segmentation literature such as capsule based segmentation~\cite{lalonde2021capsules} and transformer based segmentation~\cite{chen2021transunet}, most of the medical image segmentation literature in deep learning field are based on standard U-Net or its derivation based methodologies. In this study, we approach the segmentation from a slightly different angle where our unique clinical imaging conditions infer some constraints on the problem formulation. In many clinical scenarios, for instance, multi-modality images are necessary for a more appropriate evaluation of the clinical condition through better tissue characterization (anatomically and/or functionally). Multi-modal brain imaging, PET/CT, PET/MRI, and multi-contrast MRIs are some of the mostly used examples in this context. 

\begin{figure}[!ht]
\centering
    \includegraphics[width = 0.3\textwidth]{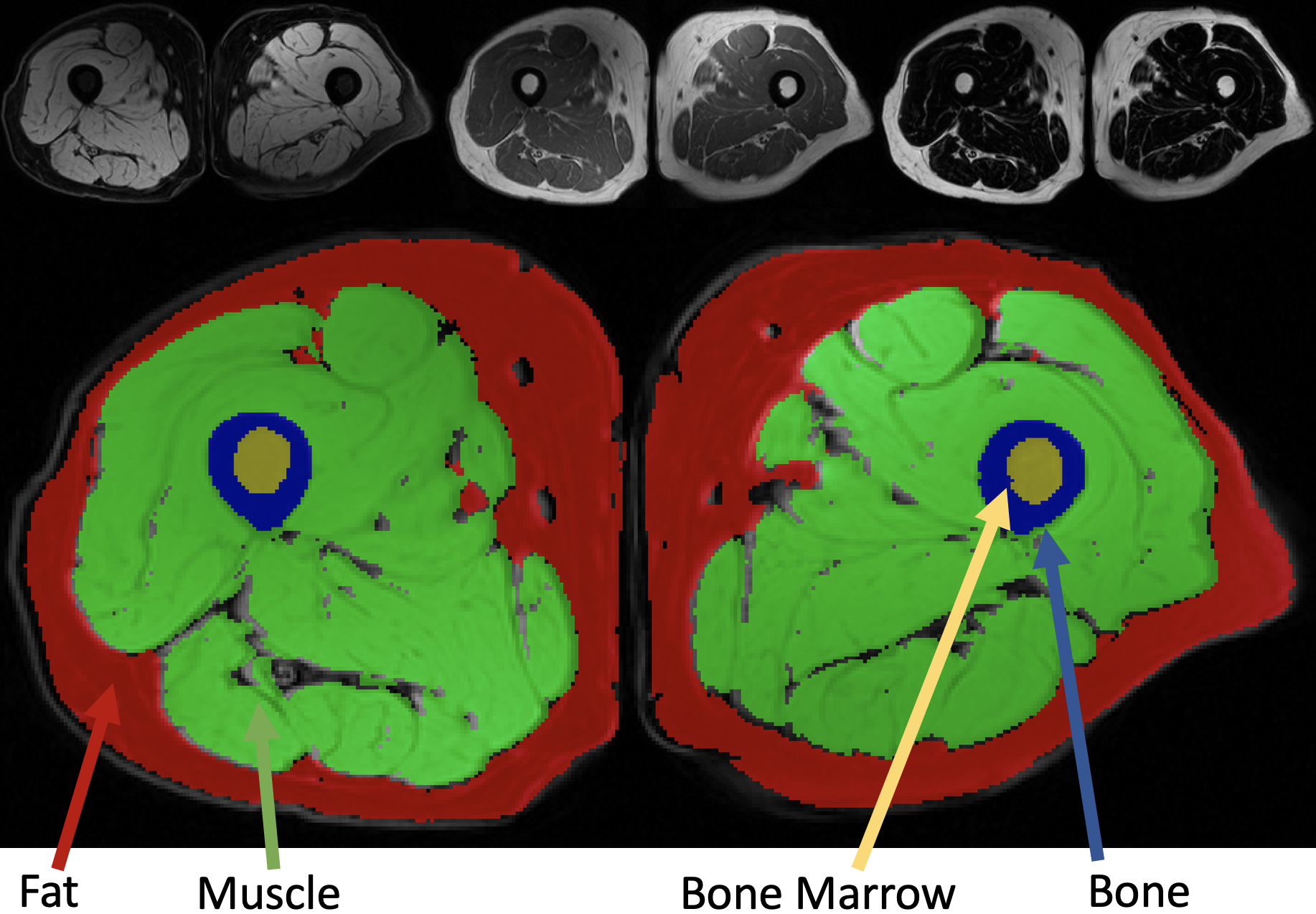}
    \caption{MRI contrasts (first row): fat-suppressed, water-fat, water-suppressed. Segmented tissues-muscle, fat, bone and bone marrow-(second row).}
    \label{fig:multi_tissue}
\end{figure}

Despite the strengths of combining multiple modality images to characterize a clinical condition better, or quantify, there are further challenges to be addressed. First, handling more than one modality for image segmentation is already more challenging than single modality images. Second, multi-object segmentation is another hurdle compared to single object segmentation, which is often the case in multi-modality image analysis. Third, clinical workflow has deficiencies and not always all modalities are available for further analysis. Missing slices or missing scans are not rare especially in multi-contrast evaluation of MRI scans. In this study, our goal is to develop a successful segmentation strategy, based on deep networks, that accepts multi-contrast MRI scans and perform a multi-tissue segmentation even there are missing scans.   

To achieve our overall goal by addressing the challenges defined above, we focus on musculoskeletal (MSK) radiology examples: delineation of thigh tissues from multi-contrast MRI scans. Figure \ref{fig:multi_tissue} demonstrates such a multi-contrast MRI scan's slices from the same patient, from left to right in top row: Fat-suppressed: MRI1, water-fat: MRI2, and water-suppressed: MRI3.  Our clinical motivation comes from the fact that MSK radiology applications are critical for several diseases spanning from obesity, metabolic syndromes to cartilage quantification. For instance, according to American Cancer Society studies in 2021 \cite{society2021cancer}, some of the most effective measures for decreasing cancer risk are having a healthy body weight, healthy diet and being physically active. Excess body weight (obesity), alcohol consumption, physical inactivity, and a poor diet are thought to be responsible for 18\% of cancer cases and 16\% of cancer deaths. Of all cancer risk factors, excess body weight is believed to be responsible for 5\% of cancers in males and 11\% of cancers in women. In this respect, sarcopenia is related with general loss of body mass  and excess body weight , and have strong relation with cancer risk factors \cite{ligibel2020sarcopenia}. 


In this work, we propose a systematic approach to (1) synthesize MRI contrasts, (2) train synthesized images on a deep learning based segmentation engine, and (3) evaluate the efficacy of the segmentation model on true multi-contrast MRI images.  We target segmenting thigh tissues (muscle, fat, bone and bone marrow). We also conduct an ablation study where training model includes true, synthesized, and mixed (true and synthesized) images towards a segmentation model. Comprehensive quantitative and qualitative segmentation results showed that proposed approach can be used effectively for multi-modal image analysis. This is essentially useful when there is not enough medical imaging data, a typical constraint in medical imaging problems. Our major contributions are as follows: 
    
\begin{itemize}
\item Application-wise our study is the first one handling missing contrast issue while retaining a high accuracy in segmenting multiple tissues from thigh MRI. 
\item Our method is generic, any deep segmentation or GAN based methods can be replaced within our framework. 
\item We present a comprehensive evaluation, carefully analyzing three contrasts of MRI, their relations and effect on the final segmentation results. 
\item We examine whether it is robust and feasible enough to run a segmentation training on completely synthesized and mixed data, opening new discussions about the use of completely synthesized data to obtain clinically accepted segmentation results on real MRI data. 
\end{itemize}

\section{Related Work}

There is a relatively small body of literature  that  is  concerned  with  muscle,  fat, bone and bone  marrow  segmentation in MSK radiology despite its clinical importance \cite{shin2021deep}. Available deep learning based studies focus on U-Net based standard segmentation methods on single or multi-tissues but mostly in single modality MRI scans. When there are missing scans, there is no particular method presented for MSK applications.

GAN (generative adversarial networks) based methods are being increasingly used for several applications spanning from brain imaging to functional imaging. One interesting work utilizes a multi-modal generative adversarial network (MM-GAN) \cite{sharma2019missing}, a variant of pix2pix \cite{isola2017image} network. Authors integrate multi-modal data from existing Brain image sequences in a single forward pass training to synthesize missing sequences. In another major work  \cite{gadermayr2019domain}, authors used popular CycleGAN network on thigh MRI for increasing the data size for segmentation purposes. Our work has some similarities with this work, but unlike focusing on data augmentation aspect of particular tissue, we  generate the whole sequence(s), using them to train segmentation models, and explore the relationship of MRI contrast in an ablation study, leading us to train the complete segmentation process on synthetic MRI scans.

In pre-deep learning era, there are some segmentation studies available too. It might be worth to mention 
that \cite{irmakci2018novel} proposed an architecture based novel affinity propagation within the fuzzy connectivity for segmenting multi contrast thigh MRI. The most recent work in this domain is handling the lack of labeling problem from a semi-supervised deep learning aspect \cite{anwar2020semi} utilizing Tiramisu network. However, the synthesis of one or more MRI contrasts remains a major challenge, and not considered in those works. Herein we propose a comprehensive evaluation and generic approach for handling missing MRI contrast and its effect on multi-tissue segmentation problems.

\begin{figure*}
\centering
    \includegraphics[width = 0.45\textwidth]{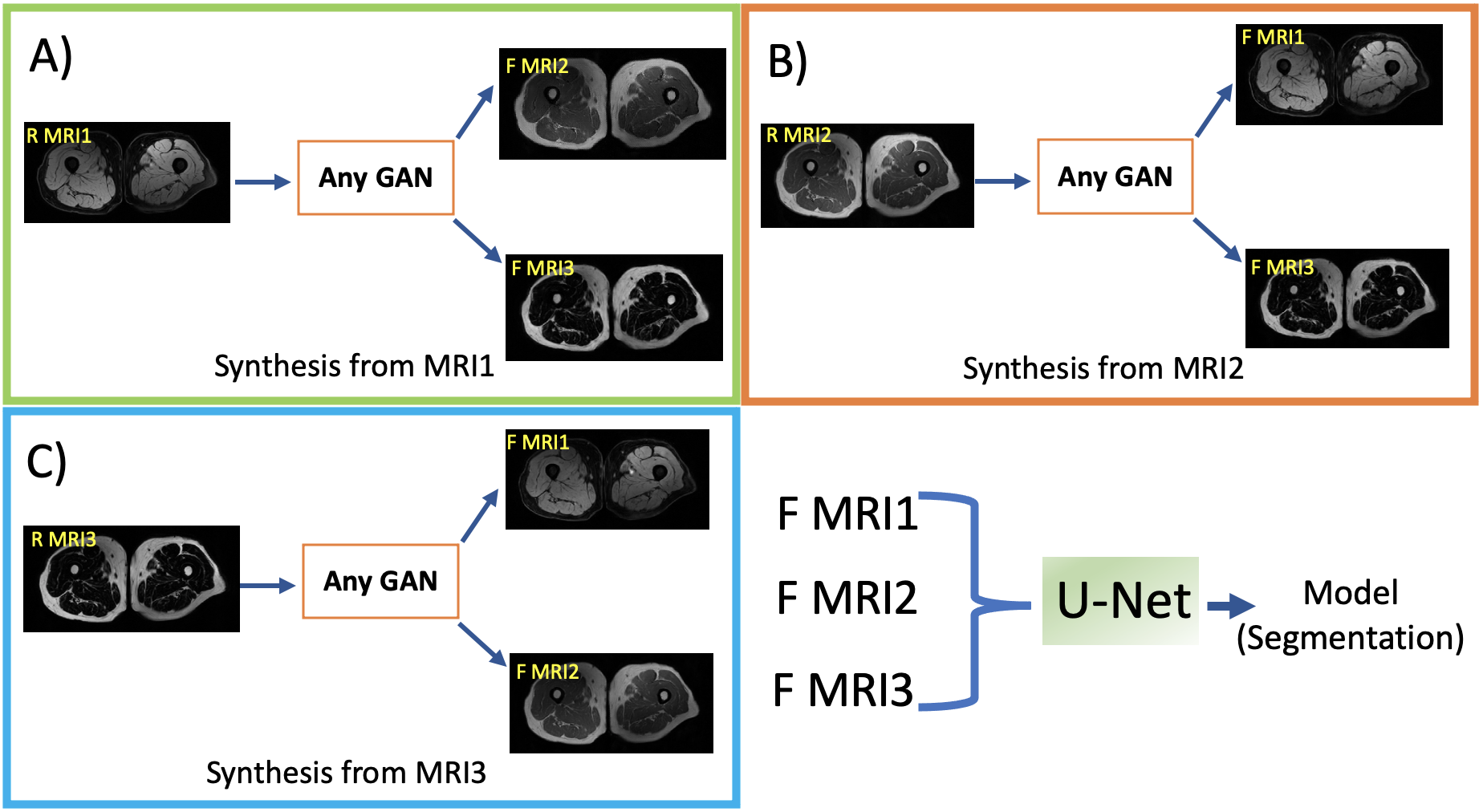}
        \includegraphics[width = 0.47\textwidth]{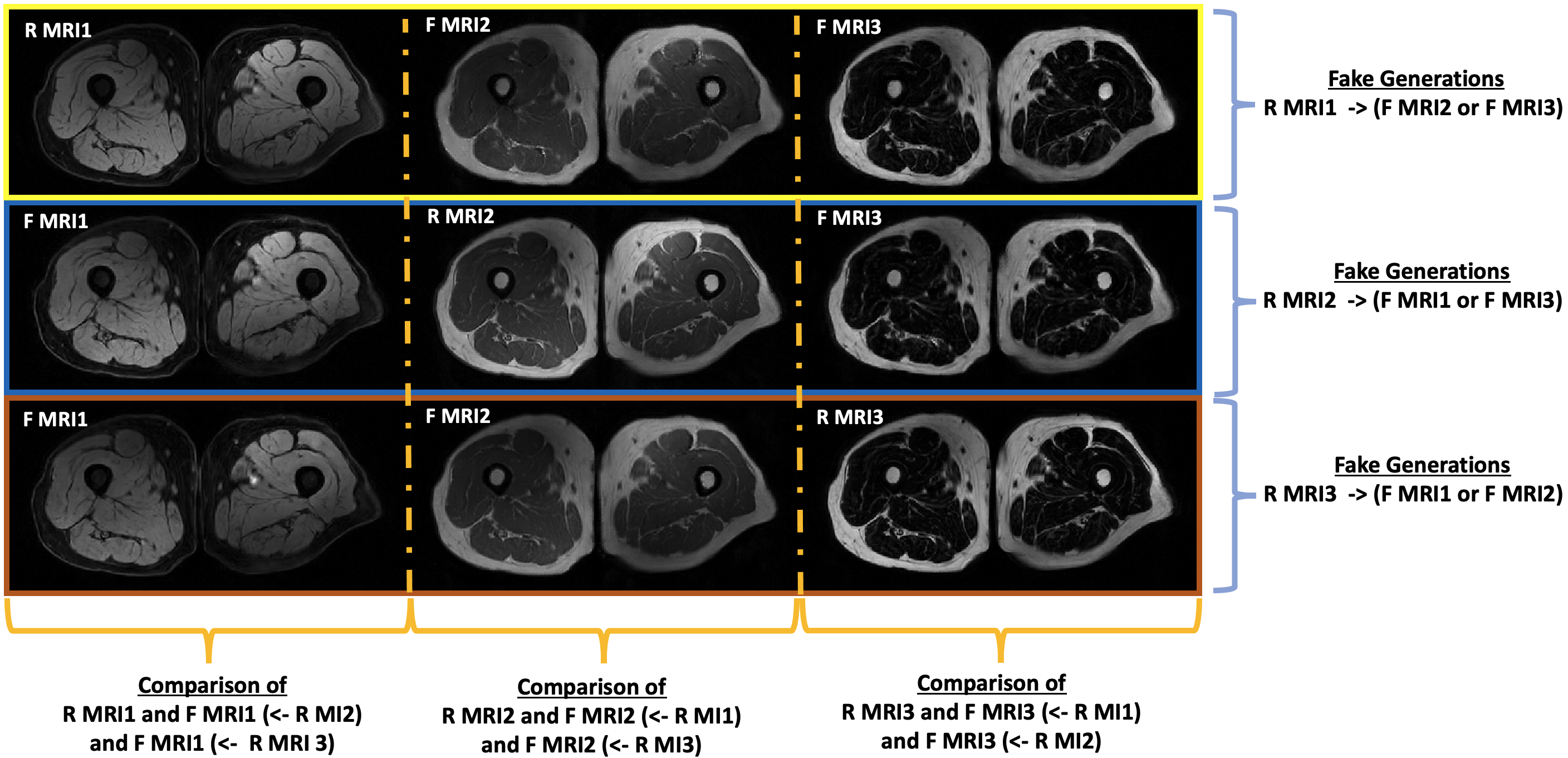}
    \caption{Fat-suppressed: MRI1, water-fat: MRI2, water-suppressed: MRI3. \textbf{Left.} Generation procedure for all Synthesized MRI contrasts. (A) Synthesized generations from only R MRI1 B) Synthesized generations from only R MRI2 C) Synthesized generations from only R MRI3. \textbf{Right.}  Different combinations of  MRI synthesis procedure are shown where ($\leftarrow$ or $\rightarrow$) indicates synthesis. For example, R MRI1$\rightarrow$F MRI2 or F MRI1$\leftarrow$R MRI2 both indicates synthesize operation from Real contrasts.}
     \label{fig:any_gan2}
\end{figure*}

\begin{table*}
\caption{Segmentation performance of Single Input Multi Output MR contrasts (5-fold cross validation) (Avg.=Average and Std.=Standard deviation) }
\centering
\resizebox{1\textwidth}{!}{%
\begin{tabular}{@{}llllllllllllllllll@{}}
\toprule
\textbf{SINGLE INPUT} & \multicolumn{5}{c}{\textbf{MUSCLE}} & \multicolumn{4}{c}{\textbf{FAT}} & \multicolumn{4}{c}{\textbf{BONE}} & \multicolumn{4}{c}{\textbf{BONE   MARROW}} \\ \midrule
 & \textbf{} & \textbf{DSC.} & \textbf{ACC.} & \textbf{SENS.} & \textbf{SPEC.} & \textbf{DSC.} & \textbf{ACC.} & \textbf{SENS.} & \textbf{SPEC.} & \textbf{DSC.} & \textbf{ACC.} & \textbf{SENS.} & \textbf{SPEC.} & \textbf{DSC.} & \textbf{ACC.} & \textbf{SENS.} & \textbf{SPEC.} \\
R  MRI1 & Avg. & \textbf{0,9264} & 0,9831 & 0,9521 & 0,9868 & \textbf{0,8826} & 0,9793 & 0,8985 & 0,9897 & \textbf{0,8245} & 0,9985 & 0,8383 & 0,9992 & \textbf{0,8397} & 0,9994 & 0,8482 & 0,9997 \\
 & Std & 0,0340 & 0,0110 & 0,0337 & 0,0097 & 0,0956 & 0,0206 & 0,0622 & 0,0126 & 0,0943 & 0,0008 & 0,0969 & 0,0005 & 0,0943 & 0,0004 & 0,1134 & 0,0002 \\
R MRI2 & Avg. & 0,9312 & 0,9846 & 0,9541 & 0,9883 & 0,9100 & 0,9870 & 0,9246 & 0,9917 & \textbf{0,9591} & 0,9997 & 0,9612 & 0,9999 & \textbf{0,9682} & 0,9999 & 0,9693 & 1,0000 \\
 & Std & 0,0384 & 0,0100 & 0,0448 & 0,0078 & 0,0811 & 0,0122 & 0,0619 & 0,0087 & \textbf{0,0321} & 0,0002 & 0,0422 & 0,0001 & \textbf{0,0254} & 0,0001 & 0,0420 & 0,0001 \\
R MRI3 & Avg. & \textbf{0,9468} & 0,9884 & 0,9587 & 0,9919 & \textbf{0,9467} & 0,9925 & 0,9608 & 0,9945 & 0,8296 & 0,9985 & 0,8324 & 0,9992 & 0,8897 & 0,9996 & 0,8848 & 0,9998 \\
 & Std & \textbf{0,0211} & 0,0057 & 0,0276 & 0,0061 & \textbf{0,0411} & 0,0057 & 0,0205 & 0,0058 & 0,0919 & 0,0009 & 0,1045 & 0,0007 & 0,0846 & 0,0003 & 0,1023 & 0,0002 \\
F MRI1($\leftarrow$ R   MRI2) TEST ON R MRI1 & Avg. & 0,9063 & 0,9774 & 0,9805 & 0,9770 & 0,8815 & 0,9817 & 0,9112 & 0,9882 & 0,8445 & 0,9986 & 0,8665 & 0,9992 & 0,8514 & 0,9994 & 0,8527 & 0,9997 \\
 & Std. & 0,0387 & 0,0130 & 0,0248 & 0,0127 & 0,0813 & 0,0157 & 0,0526 & 0,0115 & 0,0999 & 0,0008 & 0,0999 & 0,0006 & 0,0970 & 0,0004 & 0,1192 & 0,0002 \\
F MRI1($\leftarrow$ R   MRI3)  TEST ON R MRI1 & Avg. & \textbf{0,9239} & 0,9824 & 0,9550 & 0,9856 & 0,8920 & 0,9828 & 0,9301 & 0,9875 & 0,8206 & 0,9985 & 0,8263 & 0,9992 & 0,8362 & 0,9994 & 0,8372 & 0,9997 \\
 & Std. & 0,0326 & 0,0108 & 0,0334 & 0,0100 & 0,0951 & 0,0172 & 0,0537 & 0,0139 & 0,1017 & 0,0008 & 0,1021 & 0,0005 & 0,0951 & 0,0004 & 0,1168 & 0,0002 \\
F MRI2($\leftarrow$ R   MRI1) TEST ON R MRI2 & Avg. & 0,9189 & 0,9813 & 0,9661 & 0,9831 & 0,9049 & 0,9856 & 0,9468 & 0,9889 & \textbf{0,9163} & 0,9993 & 0,9387 & 0,9995 & 0,9443 & 0,9998 & 0,9550 & 0,9999 \\
 & Std. & 0,0364 & 0,0106 & 0,0309 & 0,0095 & \textbf{0,0815} & 0,0136 & 0,0539 & 0,0104 & 0,0434 & 0,0003 & 0,0403 & 0,0003 & 0,0315 & 0,0001 & 0,0521 & 0,0001 \\
F MRI2($\leftarrow$ R   MRI3) TEST ON R MRI2 & Avg. & 0,9211 & 0,9824 & 0,9422 & 0,9873 & 0,9094 & 0,9863 & 0,9312 & 0,9907 & 0,9044 & 0,9992 & 0,9061 & 0,9996 & \textbf{0,9533} & 0,9998 & 0,9547 & 0,9999 \\
 & Std. & \textbf{0,0400} & 0,0106 & 0,0491 & 0,0081 & 0,0812 & 0,0129 & 0,0567 & 0,0098 & \textbf{0,0414} & 0,0003 & 0,0552 & 0,0003 & \textbf{0,0277} & 0,0001 & 0,0532 & 0,0000 \\
F MRI3($\leftarrow$ R   MRI1) TEST ON R MRI3 & Avg. & 0,9386 & 0,9861 & 0,9622 & 0,9891 & \textbf{0,9411} & 0,9921 & 0,9754 & 0,9933 & \textbf{0,8089} & 0,9983 & 0,8496 & 0,9989 & \textbf{0,8796} & 0,9995 & 0,8785 & 0,9998 \\
 & Std. & \textbf{0,0249} & 0,0077 & 0,0323 & 0,0075 & 0,2356 & 0,2036 & 0,2067 & 0,2037 & \textbf{0,0998} & 0,0008 & 0,1052 & 0,0006 & \textbf{0,0954} & 0,0003 & 0,1075 & 0,0002 \\
F MRI3($\leftarrow$ R   MRI2) TEST ON R MRI3 & Avg. & \textbf{0,9391} & 0,9863 & 0,9677 & 0,9885 & 0,9382 & 0,9916 & 0,9709 & 0,9929 & 0,8358 & 0,9987 & 0,8370 & 0,9993 & 0,8952 & 0,9996 & 0,8891 & 0,9998 \\
 & Std. & 0,0269 & 0,0075 & 0,0239 & 0,0078 & \textbf{0,0506} & 0,0064 & 0,0165 & 0,0068 & 0,0900 & 0,0006 & 0,0937 & 0,0005 & 0,0861 & 0,0003 & 0,0983 & 0,0002 \\ \bottomrule
\end{tabular}}
\label{tab:test_on_single_mri}

\caption{Segmentation performance of Multi Input Multi Output MR contrasts (5-fold cross validation) (Avg.=Average and Std.=Standard deviation)}
\centering
\label{tab:test_on_multiple_mri}
\resizebox{1\textwidth}{!}{%
\begin{tabular}{@{}llllllllllllllllll@{}}
\toprule
\textbf{MULTI INPUT} & \multicolumn{5}{c}{\textbf{MUSCLE}} & \multicolumn{4}{c}{\textbf{FAT}} & \multicolumn{4}{c}{\textbf{BONE}} & \multicolumn{4}{c}{\textbf{BONE   MARROW}} \\ \midrule
 & \textbf{} & \textbf{DSC.} & \textbf{ACC.} & \textbf{SENS.} & \textbf{SPEC.} & \textbf{DSC.} & \textbf{ACC.} & \textbf{SENS.} & \textbf{SPEC.} & \textbf{DSC.} & \textbf{ACC.} & \textbf{SENS.} & \textbf{SPEC.} & \textbf{DSC.} & \textbf{ACC.} & \textbf{SENS.} & \textbf{SPEC.} \\
R MRI1 R MRI2 R MRI3 & Avg. & \textbf{0,9541} & 0,9898 & 0,9655 & 0,9927 & \textbf{0,9461} & 0,9923 & 0,9595 & 0,9944 & \textbf{0,9522} & 0,9996 & 0,9502 & 0,9998 & \textbf{0,9438} & 0,9998 & 0,9400 & 0,9999 \\
 & Std & 0,0200 & 0,0056 & 0,0290 & 0,0061 & 0,0419 & 0,0062 & 0,0234 & 0,0059 & 0,0410 & 0,0003 & 0,0555 & 0,0001 & 0,0617 & 0,0002 & 0,0833 & 0,0001 \\
F MRI1($\leftarrow$ R MRI2) F MRI2 ($\leftarrow$ R MRI1) F MRI3 ($\leftarrow$ R MRI1) & Avg. & 0,9365 & 0,9854 & 0,9660 & 0,9878 & 0,9357 & 0,9909 & 0,9557 & 0,9933 & 0,8795 & 0,9990 & 0,9079 & 0,9993 & 0,8928 & 0,9996 & 0,8941 & 0,9998 \\
 & Std. & 0,0280 & 0,0087 & 0,0338 & 0,0083 & 0,0520 & 0,0078 & 0,0288 & 0,0072 & 0,0862 & 0,0007 & 0,0784 & 0,0005 & 0,0868 & 0,0003 & 0,0971 & 0,0001 \\
F MRI1($\leftarrow$ R MRI2) F MRI2 ($\leftarrow$ R MRI1) F MRI3 ($\leftarrow$ R MRI2) & Avg. & 0,9370 & 0,9860 & 0,9419 & 0,9916 & 0,9341 & 0,9904 & 0,9656 & 0,9917 & 0,8913 & 0,9991 & 0,9146 & 0,9994 & 0,9145 & 0,9997 & 0,9152 & 0,9998 \\
 & Std. & 0,0327 & 0,0090 & 0,0533 & 0,0067 & 0,0557 & 0,0087 & 0,0252 & 0,0092 & 0,0841 & 0,0007 & 0,0774 & 0,0005 & 0,0794 & 0,0002 & 0,0863 & 0,0001 \\
F MRI1($\leftarrow$ R MRI2) F MRI2 ($\leftarrow$ R MRI3) F MRI3 ($\leftarrow$ R MRI1) & Avg. & 0,9281 & 0,9835 & 0,9496 & 0,9878 & 0,9249 & 0,9889 & 0,9544 & 0,9915 & 0,8960 & 0,9992 & 0,9056 & 0,9996 & 0,9276 & 0,9997 & 0,9337 & 0,9999 \\
 & Std. & 0,0306 & 0,0095 & 0,0424 & 0,0075 & 0,0618 & 0,0105 & 0,0363 & 0,0089 & 0,0803 & 0,0006 & 0,0788 & 0,0004 & 0,0700 & 0,0002 & 0,0850 & 0,0001 \\
F MRI1($\leftarrow$ R MRI2) F MRI2 ($\leftarrow$ R MRI3) F MRI3 ($\leftarrow$ R MRI2) & Avg. & \textbf{0,9408} & 0,9868 & 0,9549 & 0,9908 & 0,9314 & 0,9900 & 0,9558 & 0,9924 & 0,9011 & 0,9992 & 0,9108 & 0,9996 & 0,9240 & 0,9997 & 0,9303 & 0,9999 \\
 & Std. & 0,0292 & 0,0081 & 0,0430 & 0,0066 & 0,0582 & 0,0093 & 0,0374 & 0,0078 & 0,0765 & 0,0006 & 0,0727 & 0,0004 & 0,0773 & 0,0002 & 0,0858 & 0,0001 \\
F MRI1($\leftarrow$ R MRI3) F MRI2 ($\leftarrow$ R MRI1)  F MRI3 ($\leftarrow$ R MRI1) & Avg. & 0,9375 & 0,9860 & 0,9526 & 0,9902 & 0,9346 & 0,9909 & 0,9552 & 0,9930 & 0,8879 & 0,9991 & 0,8919 & 0,9995 & 0,8934 & 0,9997 & 0,8876 & 0,9999 \\
 & Std. & 0,0263 & 0,0076 & 0,0395 & 0,0069 & \textbf{0,0515} & 0,0072 & 0,0295 & 0,0073 & 0,0788 & 0,0006 & 0,0772 & 0,0004 & 0,0887 & 0,0003 & 0,1073 & 0,0001 \\
F MRI1($\leftarrow$ R MRI3) F MRI2 ($\leftarrow$ R MRI1)  F MRI3 ($\leftarrow$ R MRI2) & Avg. & 0,9406 & 0,9868 & 0,9518 & 0,9911 & 0,9349 & 0,9908 & 0,9680 & 0,9919 & 0,8925 & 0,9991 & 0,8990 & 0,9995 & 0,9127 & 0,9997 & 0,9069 & 0,9999 \\
 & Std. & \textbf{0,0249} & 0,0071 & 0,0412 & 0,0062 & 0,0515 & 0,0075 & 0,0257 & 0,0078 & 0,0693 & 0,0005 & 0,0734 & 0,0003 & 0,0757 & 0,0003 & 0,0857 & 0,0001 \\
F MRI1($\leftarrow$ R MRI3) F MRI2 ($\leftarrow$ R MRI3)  F MRI3 ($\leftarrow$ R MRI1) & Avg. & 0,9288 & 0,9843 & 0,9344 & 0,9906 & 0,9254 & 0,9888 & 0,9489 & 0,9913 & \textbf{0,9061} & 0,9992 & 0,9203 & 0,9996 & \textbf{0,9375} & 0,9998 & 0,9295 & 0,9999 \\
 & Std. & 0,0336 & 0,0094 & 0,0519 & 0,0064 & 0,0624 & 0,0108 & 0,0421 & 0,0096 & \textbf{0,0570} & 0,0004 & 0,0586 & 0,0003 & \textbf{0,0526} & 0,0002 & 0,0737 & 0,0001 \\
F MRI1($\leftarrow$ R MRI3) F MRI2 ($\leftarrow$ R MRI3)  F MRI3 ($\leftarrow$ R MRI2) & Avg. & 0,9399 & 0,9865 & 0,9519 & 0,9909 & 0,9335 & 0,9906 & 0,9611 & 0,9924 & 0,8840 & 0,9991 & 0,8808 & 0,9996 & 0,9180 & 0,9997 & 0,9193 & 0,9999 \\
 & Std. & 0,0256 & 0,0074 & 0,0384 & 0,0063 & 0,0537 & 0,0077 & 0,0275 & 0,0075 & 0,0794 & 0,0005 & 0,0892 & 0,0003 & 0,0756 & 0,0002 & 0,0882 & 0,0001 \\ \bottomrule
\end{tabular}}
\caption{Quality of Single Modality Thigh MRIs Synthesis with 5-fold cross validation. (Avg. = Average (Mean) and Std.=Standard deviation)}
\label{tab:psnr_fid_ssim}
\resizebox{1\textwidth}{!}{
\begin{tabular}{@{}llllllll@{}}
\toprule
 &  & F  MRI1($\leftarrow$ R MRI2) & F  MRI1($\leftarrow$ R MRI3) & F  MRI2($\leftarrow$ R MRI1) & F  MRI2($\leftarrow$ R MRI3) & F  MRI3($\leftarrow$ R MRI1) & F   MRI3($\leftarrow$ R MRI2) \\ \midrule
PSNR & Avg. & 28,3153 & 27,6520 & 27,2156 & 27,9233 & \textbf{28,5335} & 28,0810 \\
 & Std. & 3,3855 & \textbf{2,9039} & 3,2992 & 3,4147 & 3,2828 & 3,7948 \\
SSIM & Avg. & 0,8786 & 0,8848 & 0,8728 & 0,8827 & \textbf{0,8968} & 0,8890 \\
 & Std. & \textbf{0,0496} & 0,0510 & 0,0520 & 0,0608 & 0,0601 & 0,0616 \\
FID & Avg. & 42,5333 & 41,6491 & 57,5200 & \textbf{68,3417} & 39,8365 & 54,2351 \\
 & Std. & \textbf{4,4439} & 4,6984 & 10,9396 & 20,8823 & 4,4900 & 13,9198 \\ \bottomrule
\end{tabular}}
\end{table*}

\section{METHOD}

The proposed segmentation strategy includes both synthesis of missing contrasts with a generator and a segmentor (Figure~\ref{fig:any_gan2}). In the generation stage, we adapted popular pix2pix \cite{isola2017image} conditional GAN method for synthesising all synthesized contrasts from real (true) contrasts. Any other GAN method can be used too. Briefly, pix2pix uses a conditional generative adversarial network to learn mapping from a source domain $x$ and random noise $z$ to an target domain $y$. The network is made up of two  blocks, the generator ${G}$, and the discriminator ${D}$. The generator transforms the source domain ($x$) with random noise ($z$) to get the target domain ($y$) while discriminator learns how much target domain is similar to source domain. As shown in Figure ~\ref{fig:any_gan2}, then we use all real MR contrasts (R MRI) for synthesizing (generating) other contrasts (F MRI)  using Equation \ref{eq:1} and Equation \ref{eq:2} where $x$ is source contrast, $y$ is target contrast, $z$ is random noise, $\lambda$ is hyperparameter for adjusting blurriness, and $\mathcal{L}_{L 1}$ mean square error: 

\begin{equation}
\begin{aligned}
\label{eq:1}
\mathcal{L}_{c G A N}(G, D)=& \mathbb{E}_{x, y}[\log D(x, y)]+\\
& \mathbb{E}_{x, z}[\log (1-D(x, G(x, z))].
\end{aligned}
\end{equation}

\begin{equation}
\label{eq:2}
G^{*}=\arg \min _{G} \max _{D} \mathcal{L}_{c G A N}(G, D)+\lambda \mathcal{L}_{L 1}(G).
\end{equation}

First, we condition on source contrast, Real MRI1 for generating Synthesized MRI2 (R MRI1 $\rightarrow$ F MRI2) or Synthesized MRI3 (R MRI1 $\rightarrow$ F MRI3) separately, then we condition on Real MRI2 for generating Synthesized MRI1 (R MRI2 $\rightarrow$ F MRI1) or Synthesized MRI3 (R MRI2 $\rightarrow$ F MRI3), separately. Finally, we condition on Real MRI3 for generating Synthesized MRI1 ((R MRI3 $\rightarrow$ F MRI1) and Synthesized MRI2 (R MRI3 $\rightarrow$ F MRI2) for getting all six different combinations.



For delineation of multiple tissues, we devise  commonly used standard U-Net segmentor \cite{ronneberger2015u}. We speculate that if synthesized images are in good quality (Table \ref{tab:psnr_fid_ssim}), overall segmentation of tissues should be accurate. In addition to fully synthesized, mixed, and real images based training of segmentation, we also apply multi-contrast and single-contrast segmentation settings to validate the necessity of additional contrasts and their complementary strengths to each other.


\section{EXPERIMENTS and RESULTS}

\label{sec:typestyle}
\noindent\textbf{Dataset and Preprocessing:} We have used multi-contrast MRI data from Baltimore Longitudinal Study of Aging (BLSA) \cite{ferrucci2008baltimore}. Experiments were performed on three different T1-weighted MR contrasts:  fat-suppressed  (MRI1), water and fat (MRI2),and water-suppressed (MRI3). These images are abbreviated as "real" in our experiments to separate them from synthesized ones. Original data set contains 150 volumetric MRI scans from 50 subjects acquired using a 3T Philips Achieva MRI scanner (Philips Healthcare, Best, The Netherlands). A voxel size of $1 \times 1$ mm$^2$ in-plane, and slice thickness varies from 1 mm to 3 mm in different scans. Details of contrasts and other imaging parameters can be found in \cite{ferrucci2008baltimore}. Prior to our experiments, we used non-uniform non-parametric  intensity  normalization  technique (N4ITK) \cite{tustison2010n4itk}  to  remove  bias  afterwards  edge-preserving  diffusive filter  for  removing  noise  without any distortion on tissue structures in  MR  images. Finally, we used whitening transformation then scaled voxel values between $0$ and $1$.\\
 
 \noindent\textbf{Network settings and Training Procedure:} pix2pix was trained for 250 epoch based on 2D slices with learning rate of 0.0001. Generator and discriminators consist of U-Net and PatchGAN. Best models were selected on validation portion of the whole data set. In segmentation stage, we optimized network with cross entropy loss with ADAM optimizer and a learning rate of 0.0001. Early stopping criteria was used. We did not use any data augmentation techniques as we did not see any overfitting problem in training. We have  performed 90 (18x5 Fold) experiments in segmentation stage and 30 (6x5 Fold) experiments for generation stage with 5 fold cross validation ($70\%$ training, $10\%$ validation, and $20\%$ test). All experiments were performed on Nvidia Titan-XP GPUs with 12GB memory.  Proposed  approach  was  implemented  in  PyTorch framework.\\
 
\noindent\textbf{Quantitative evaluations:} We report our GAN results in three different metrics: the Frechet Inception Distance (FID, lower is better), Peak Signal to Noise Ratio (PSNR, higher is better) and Structural Similarity Index Measure (SSIM, higher is better) (Table \ref{tab:psnr_fid_ssim}). We observed that synthesized MRI3 from real MRI1 gives the best PSNR,  SSIM and FID outperforming other synthesized MRI contrasts. However, results are not hugely different from each other, this is likely due to the one-to-one mapping nature between MRI contrasts and the learned maps between contrasts are highly informative about each other. This is a strength that can be attributed to the power of GAN.

We also summarized segmentation results with evaluation metrics of DICE, accuracy, sensitivity, and specificity (Table \ref{tab:test_on_single_mri} and Table \ref{tab:test_on_multiple_mri}). We analyzed the muscle, fat, bone and bone marrow segmentation results on single synthesized and real MRI contrasts (Table \ref{tab:test_on_single_mri}). Muscle and fat tissue shows higher DICE scores  when real MRI3 (water-suppressed) is used, and bone and bone marrow show higher DICE scores when R MRI2 (water-fat) is used. Surprisingly, synthesized MRI3 ($\leftarrow$ R MRI2), synthesized MRI3 ($\leftarrow$ R MRI1), synthesized MRI2 ($\leftarrow$ R MRI1), synthesized MRI2 ($\leftarrow$ R MRI3) show similar results for muscle, fat, bone and bone marrow tissues, respectively (Table \ref{tab:test_on_single_mri}), thanks to strongly learned mapping between each contrast. For multi-contrast input (Table \ref{tab:test_on_multiple_mri}), although true MRIs trained segmentation results show higher DICE scores than other strategies, synthesized images from real MRI2 and MRI3 show very close DICE scores to that of the best results. Synthesized images based segmentation results showed similar trends to true images based segmentation results even when the tissue of interest is small such as bone marrow, indicating that high-quality synthesis was achieved.\\

\begin{figure}[!ht]
\centering
    \includegraphics[width = 0.47\textwidth]{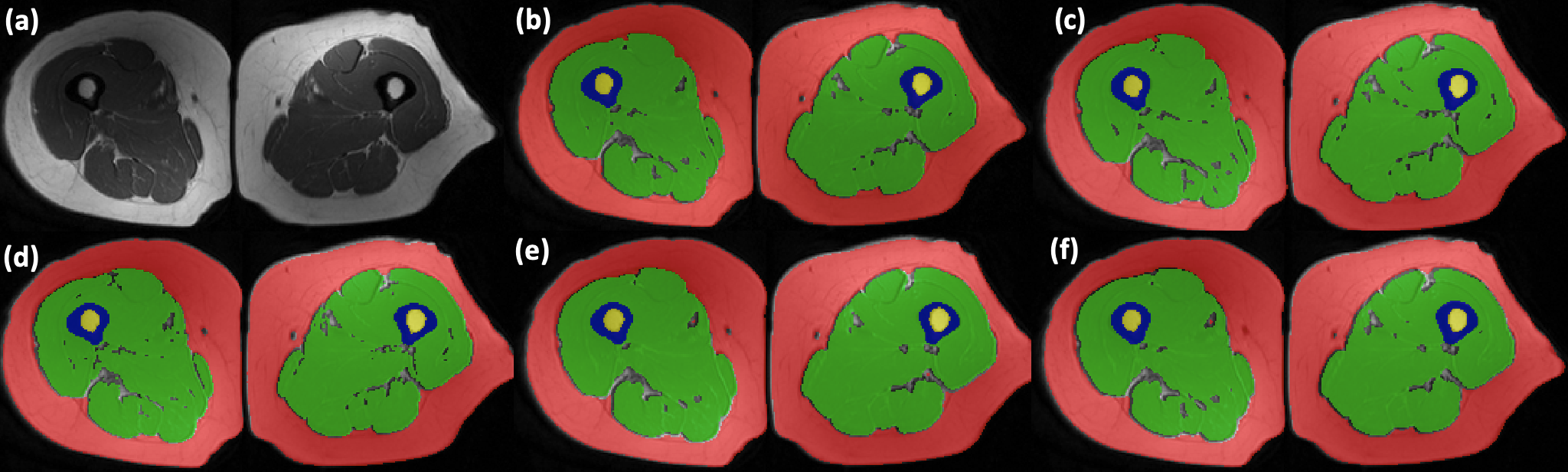}
    \caption{Comparison of fat, muscle, bone and bone marrow tissue segmentation results for a given MRI slice. (a) Original MRI, b) F MRI1($\leftarrow$ R MRI2) F MRI2 ($\leftarrow$ R MRI3) F MRI3 ($\leftarrow$ R MRI2) c) F MRI1($\leftarrow$ R MRI3) F MRI2 ($\leftarrow$ R MRI3) F MRI3 ($\leftarrow$ R MRI2) d) R MRI1 R MRI2 R MRI3 e) F MRI2 ($\leftarrow$ R MRI1) f) F MRI2 ($\leftarrow$ R MRI3).}
    \label{fig:qualitative}
\end{figure}

\noindent\textbf{Qualitative evaluations:} Qualitative results are shown in Figure \ref{fig:qualitative} for muscle, fat, bone and bone marrow tissues. We compare some best synthesized MRI image results (Table \ref{tab:test_on_single_mri} and Table \ref{tab:test_on_multiple_mri}) to original MRIs. We observed that synthesizing multi-contrasts MRI were high quality such that even small details on soft tissues were preserved, justified in segmentation results too. This observation is promising as lack of data, missing contrast and other data problems can be addressed with synthetic images as an intermediate step while diagnostic path is still using true/real images, thus avoiding the concern of using synthetic images as diagnostic purpose.

\section{CONCLUSIONS}

\label{sec:majhead}
 In this work, we conducted extensive experiments to explore the use of synthetic MRI scans for training segmentation engine for multi-tissue analysis. 
 We showed that an accurate segmentation model can be built on solely based on synthetic scans or mixed (real + synthetic) images with a precision level close to a segmentation model completely trained with true images. In addition, we have demonstrated that multi-modality combination of scans provide better segmentation results even when some of the modalities are synthetically generated. \\
 
\noindent\textbf{Acknowledgments}
Our study is exempt from human subject section as the data is publicly available and fully anonymized. This study is approved under the existing IRB at Baltimore Longitudinal Study of Aging (BLSA) \cite{ferrucci2008baltimore}. This study is partially supported by the NIH grant R01-CA246704-01 and R01-CA240639-01. We thank Ege University for letting us to use their servers for running our experiments. 
\addtolength{\textheight}{-12cm}   

\bibliographystyle{IEEEbib}
\bibliography{refs}

\end{document}